# New type of thermoelectric conversion of energy by semiconducting liquid anisotropic media

## S.I. Trashkeev,* A.N. Kudryavtsev **

*Institute of Laser Physics, Siberian Branch of Russian Academy of Sciences (Novosibirsk)*
***Khristianovich Institute of Theoretical and Applied Mechanics,*
*Siberian Branch of Russian Academy of Sciences (Novosibirsk)*
*e-mails: sitrskv@mail.ru; alex@itam.nsc.ru*

## Abstract

The paper describes preliminary investigations of a new effect in conducting anisotropic liquids, which leads to thermoelectric conversion of energy. Nematic liquid crystals with semiconducting dopes are used. A thermoelectric figure of merit $ZT \approx 0.2$ is obtained in experiments. The effect can be explained by assuming that the thermocurrent in semiconducting nematics, in contrast to the Seebeck effect, is a nonlinear function of the temperature gradient and of the temperature itself. Though the discovered effect has to be further investigated, the data obtained suggest that it can be effectively used in alternative energy engineering.

## 1. INTRODUCTION

The topic of this paper refers to problems of alternative energy engineering or the so-called unconventional methods of electric energy production [1]. There is no need to prove the importance of using environmentally friendly renewable sources of energy. Many specialists believe that the mere existence of humankind depends on solving these problems.

The most widespread type of energy is thermal energy, whereas the most universal type from the viewpoint of usability by humans is electricity. If we assume that all types of energy can be transformed to heat almost without any losses, then we see that direct (without intermediate conversion of energy to a third type) and highly efficient conversion of heat to electricity (thermoelectric conversion, TE conversion) can solve many problems of power engineering. The majority of activities in the field of alternative energy engineering are aimed either at searching for new sources of energy or at improving available methods and materials for energy conversion [1, 2]. In the thermoelectric aspect, one of the main methods of direct energy conversion is the Seebeck thermocouple effect discovered in 1821. Significant progress has been achieved by now [2] in applications of the Seebeck effect in semiconductor materials [3], but thermoelectric systems developed on the basis of this effect do not satisfy some parameters required for their application and commercialization. First of all, the converters have a low efficiency and (or) are too expensive to be widely used.

The paper describes research results that lead to a conclusion about the existence of a new principle of thermoelectric conversion discovered in anisotropic semiconducting organic liquids. Mixtures of liquid crystals (LCs) with dopes imparting semiconductor properties to the medium were used as such a medium. The observed phenomenon can be considered as a nonlinear Benedicks effect. The Benedicks effect [4] was earlier considered to be rather small as compared to the Seebeck effect and was observed only under extreme conditions [5]. The results reported in this paper cannot be regarded as record-breaking ones, such as those obtained on the basis of the Seebeck effect in bismuth telluride (Bi$_2$Te$_3$) [2], but the efficiency and the electric power obtained in these experiments are higher by more than an order of magnitude than those of organic thermoelectric converters based on carbon nanotubes [6]. We think that discovery of a new TE conversion type in soft matters is an interesting scientific result, which can contribute to physics of condensed matter [7]. Though the theory of the process has not be finalized yet, the estimates show that anisotropic liquids and this effect can provide a possibility of TE conversion with the efficiency close to that of the Carnot cycle. In our opinion, this effect can be essential for solving the problem of using alternative renewable sources of energy.

## 2. EXPERIMENTAL DATA

To study the thermoelectric effect, we prepared samples with a layered structure. One layer was limited on both sides by solid (rigid or flexible) substrates, and the gap between them was filled by an anisotropic liquid (working substance). The working substances were pure LCs or their mixtures, and also LC mixtures with dopes that give a semiconductor state with a higher conductivity than that of original LCs. For comparison, control experiments were performed with usual isotropic liquids, where this effect was not observed. The substrates simultaneously served as electrodes on which the voltage was measured. The gap between the electrodes was defined by non-continuous shims with a calibrated thickness, which were mounted at the edges of the rigid substrates; or the entire gap between the substrates (electrodes) was filled by a porous film impregnated by the working liquid. To eliminate (minimize) the contribution of galvanic effects to converter operation, the electrodes were made of chemically identical substances. The surface of one electrode was treated to create a certain profile. The shape of the surface profile, the methods of surface treatment, and the compositions and proportions of the working liquids used in these experiments can be patented in the future. The new TE conversion was first claimed in a patent obtained in 2001 [8], where this phenomenon, however, was erroneously attributed to the Seebeck effect.

Stainless steels, carbon (C), aluminum (Al), copper (Cu), titanium (Ti), titanium nitride (TiN), nickel (Ni), and ITO (In$_2$O$_3$ + SnO) were considered as electrode (contact) materials. Among the working liquids and surface treatment methods of our experiments, practically no thermoelectric effect was observed for carbon, nickel, and titanium, whereas the greatest effect was obtained for Al and TiN. Possibly, there are some other anisotropic liquids, electrode materials, and surface treatment methods that can ensure a higher TE conversion efficiency that that obtained in this work. Despite a limited number of materials and technologies tested, a preliminary conclusion was drawn that the electrode material should possess semiconductor properties, like the TiN and ITO compounds used in these experiments and also oxides formed on Al and Cu surfaces.

A multi-layered TE converter sample is schematically shown in Fig. 1. Electrode groups 1 and 2 were made of chemically identical materials. The neighboring electrodes differed only by surface treatment, which we tried to perform with the minimum possible contamination by byproducts. After shaping of some metal surfaces, the oxide layer (especially for Al) became appreciably thicker. The gap between the electrodes was filled by the working medium 3, which could be of two types. In one-layer samples with rigid substrates fixed with respect to each other by shims, the gap was filled by the working liquid. Metal plates or ITO glasses could be used as substrates-electrodes. In multi-layered samples with thin flexible electrodes (metal foils), porous inserts impregnated by the working liquid were used. Such structures allowed us to generate a nonlinear (with a variable gradient, see the wavy line in Fig. 1) distribution of temperature at $T_1 \neq T_2$ inside the sample.

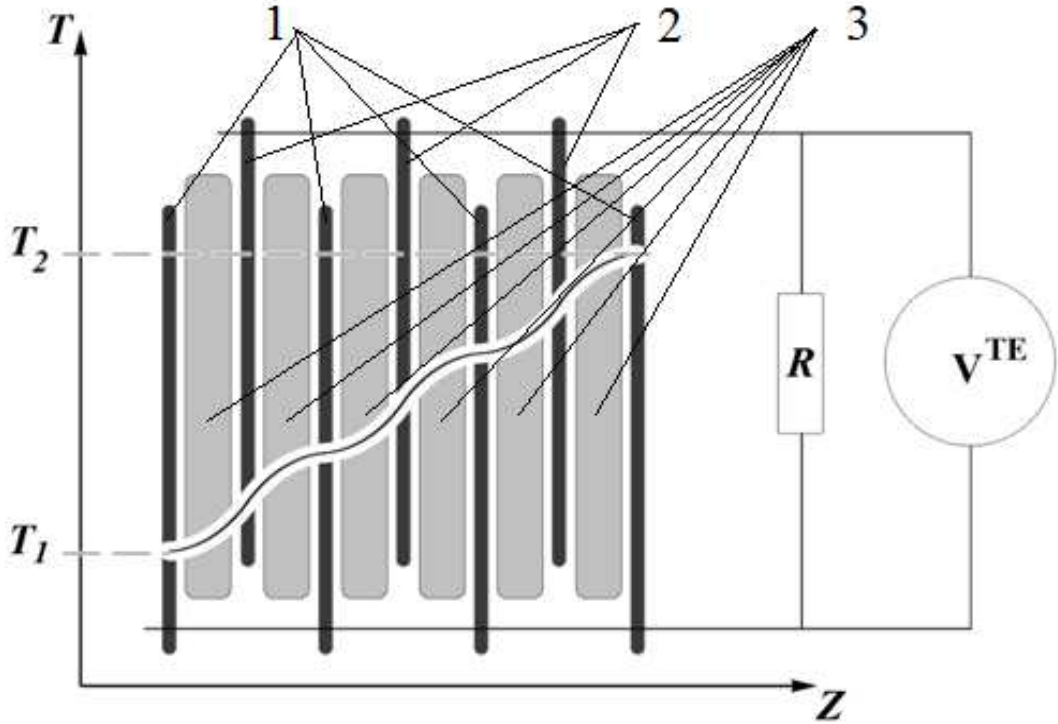

Fig.1.Sample of multi-layered TE converter. The electrodes 1 and 2 are made of chemically identical conducting materials. The surfaces of the group of electrodes 1 are contoured and the surface shape differs from that of the groups of electrodes 2. The working substance is indicated by 3. The wavy line is the temperature profile, $T = T(z)$, $R$ is the loading resistance, and $V^{TE}$ is the measured thermoelectric voltage.

It was found that the electric parameters of the converter depend in a complicated (nonlinear) and often non-unique manner not only on the difference in the electrode temperatures, but also on the temperatures themselves. Heating or cooling dynamics exerts an explicit effect. An unsteady behavior of the dependences of the thermal voltage and thermal current on time can be observed at steady temperatures. Additional uncertainty in systematization of experimental data is induced by the supercapacitor property (the effective capacity can reach several farads) and the dependence of the electric parameters of the converter on the load or current in the circuit. It does not seem possible to describe and systematize all experimental data in one publication; it is more convenient to do it later, when a completed theoretical model of the TE conversion is constructed. We intend merely to state the fact of discovery of a new phenomenon and give only some typical steady electrotechnical indicators and dependences among the experimental results.

First of all, the most dramatic difference between the TE conversion and the Seebeck effect should be noted. The sign of the thermal electric current is independent of the fact which surface is heated or cooled; this sign is determined by the type of processing of the contact surfaces.

Figure 2 shows the dependence of the thermovoltage $V$ (thermo-EMF) of a one-layer converter without the load ($R \rightarrow \infty$) on the mean temperature for the temperature difference $\Delta T \approx 1$°C. The examined sample was made of aluminum disks 60 mm in diameter, the gap between the disks was 5 μm, and a pure undoped nematic liquid crystal (NLC) consisting of a mixture of cyanobiphenyls was used. The mean thermal conductivity was $\bar{\lambda}_{LC} \approx 0.25$ W$\cdot m^{-1} K^{-1}$. The internal resistance of the sample depended on temperature and was 3 MΩ or more. Oxide layers of the electrodes contributed to the sample resistance. As is seen from Fig. 2, the thermoelectric parameters of conversion strongly depend on the mean temperature, which is typical for the majority of the properties of LC media [9]. The maximum value of the thermal voltage reached is $V \approx 0.8$ V.

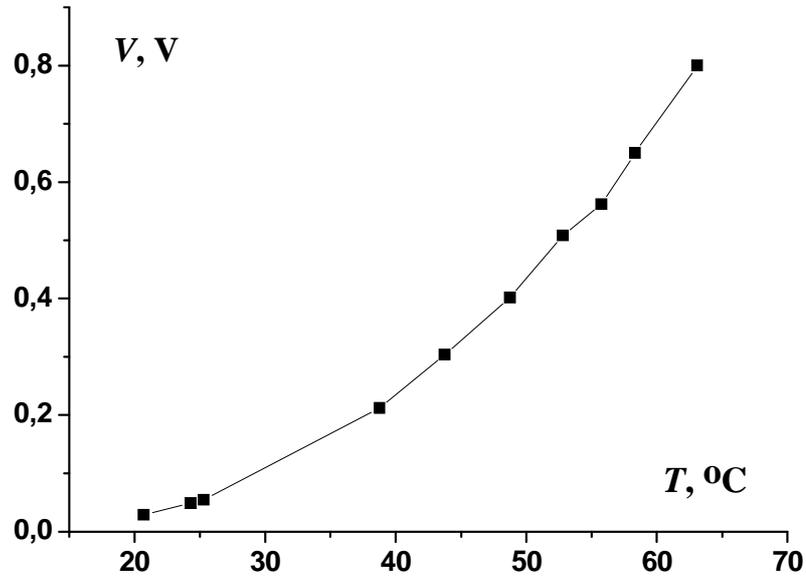

Fig. 2. Thermo-EMF versus the mean temperature in a one-layer sample.

The dependence of thermo-EMF on the temperature difference is nonlinear. At small temperature differences (up to $\Delta T \sim 1°$), the thermovoltage grows, then, with increasing $\Delta T$, it saturates or starts to drop. The behavior depends on the type of material and both the working temperatures. Such a dependence, as well the constant direction of thermocurrent (regardless of the sign of $\Delta T$), disagree with the linear Seebeck law. In all performed experiments, optimal, for the used materials, values of the temperature difference in the range of $\Delta T_{опт} \approx 0.5 - 3\ °C$ were chosen, which provided the maximum electrical power at the minimum heat flux. The resulting value $\Delta T_{опт}$ is a consequence of nonlinearity of the investigated TE conversion that limits the thermovoltage. Reasons for existence of the optimal temperature difference are considered in more detail in the next section. This disadvantage can be removed by selecting other materials or utilizing multilayer constructions.

The internal resistance of the samples based on pure NLCs whose conductivity was $\sigma \sim 10^{-8} - 10^{-9}\ \Omega^{-1}\cdot m^{-1}$ [9, 10] did not allow us to obtain high values of the thermocurrent. Semiconducting alloying dopes were used to increase the internal conductivity.

Figures 3-5 show the electric parameters obtained for a 19-layer sample with aluminum electrodes, which is schematically shown in Fig. 1. The porous inserts 3 were impregnated by cyanobiphenyls with a semiconducting dope. The thickness of one porous insert was 25 μm, and the electrode diameter was 55 mm (the electrodes were made of aluminum foil). The total thickness of the sample was 1.3 mm. The mean thermal conductivity of the insert impregnated by the NLC was $\bar{\lambda} \approx 0.12\ W\cdot m^{-1}K^{-1}$. When porous inserts and doped NLCs were used, the thermo-EMF was lower than that in samples prepared from pure NLCs (see Fig. 2). In the 19-layer sample, the dependence of the thermo-EMF on the mean temperature had a similar form and EMF reached 0.68 V (at 93°C).

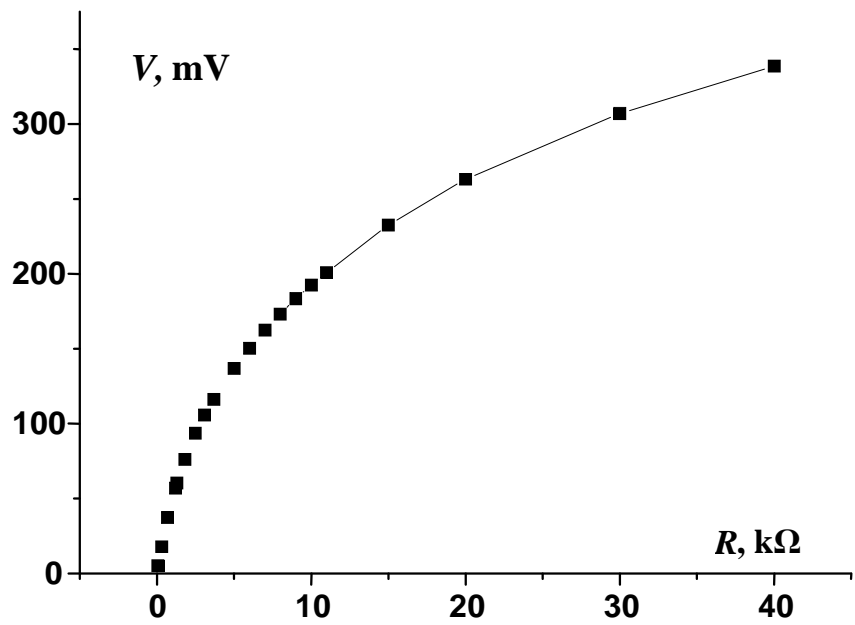

Fig. 3. Thermovoltage versus the load at $T_1 = 95°C$ and $T_2 = 96°C$.

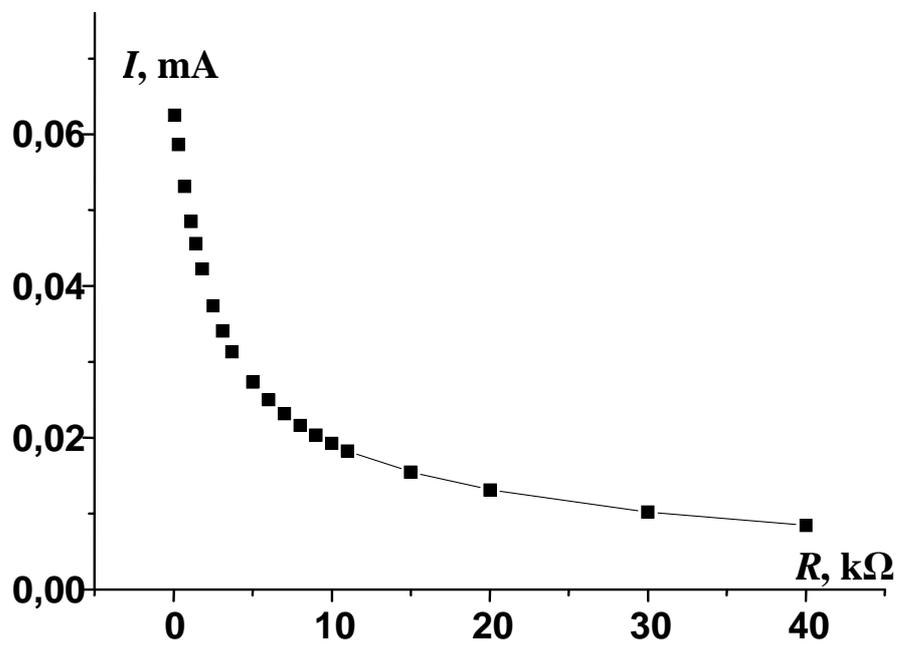

Fig. 4. Thermocurrent versus the load at $T_1 = 95°C$ and $T_2 = 96°C$.

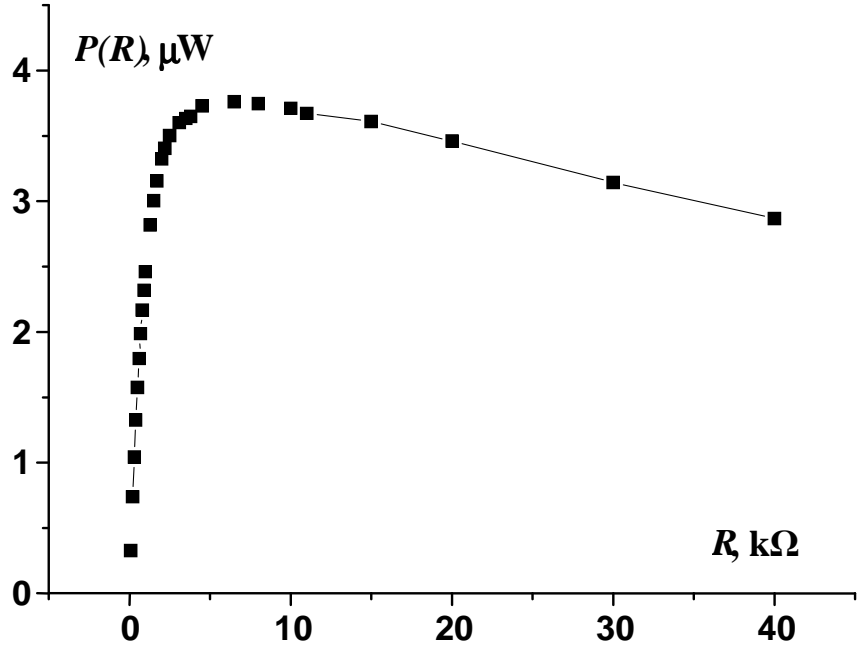

Fig. 5. Power released on the load versus the load resistance at $T_1 = 95°C$ and $T_2 = 96°C$.

Figures 3 and 4 show the thermovoltage $V$ and the thermocurrent $I$ as functions of the load resistance $R$. It is seen from Fig. 5, which shows the power $P$ released on the load $R$ versus the value of $R$, that the maximum power equal to 3.78 µW is reached at $R = 5.8$ kΩ. After equalization of temperatures ($T_1 = T_2$), the voltage and current slowly decrease to zero (capacitor discharging occurs). Based on the discharging time, it is possible to estimate the element capacity, which reached 2.7 F at $T_1 = T_2 = 96°C$ and R = 5.8 kΩ.

Using the data of Fig. 5 and the heat and power parameters of the converter used, we can estimate the effective value of the thermoelectric figure of merit $(ZT)_{LC}$ by defining the latter as the efficiency divided by the temperature difference and multiplied by the mean temperature [2,3],

$$(ZT)_{LC} = \eta \left( \frac{T_{av}}{\Delta T} \right),$$ (1)

where the efficiency $\eta = P_E / P_T$ is the ratio of the electric power to the heat flux power. In our case, we have $(ZT)_{LC} \approx 0.25$. For comparison, the same quantity for the Power Felt material was estimated in [6] as $(ZT)_{PF} = \alpha^2 \sigma T_{av} / \lambda \approx 0.02$.

At this stage of research, we did not perform detailed investigations of the influence of various possible chemical (galvanic) processes in the working medium on the TE conversion, though we did our best to minimize them. It is known [9, 11] that LCs always contain small amounts of ionic species, and reactions due to interactions between the dissolved oxygen, electrodes, LCs, and alloying dopes can occur. The influence of polarization effects associated with flexoelectric and pyroelectric effects is also possible. All these effects can lead to undesirable changes in the thermoelectric parameters. In further studies, the chemical and other processes inherent in the TE converter will be considered in more detail and eliminated to the maximum possible extent.

## 3. POSSIBLE THEORETICAL EXPLANATION. PHENOMENOLOGICAL APPROACH

Operation of thermal machines converting heat to other types of energy is based on the dependence of the working medium energy on its temperature. It was shown in [12, 13], where the thermal orientation effect was discovered, that the free energy of the NLC depends not only on the temperature proper, but also on the temperature gradient. The dependence was determined by an additional term of the NLC free energy $F_{te}$ in a quadratic form of the temperature gradients

$$F_{te} = \frac{1}{2} \frac{\partial T}{\partial x_i} \gamma_{ij} \frac{\partial T}{\partial x_j}, \quad x_i, x_j = x, y, z, \tag{2}$$

where $\gamma_{ij} = \eta \lambda_{ij}$ is the thermal tensor introduced and measured in [13], whose mean value is $\sim 10^{-11}$ N/K$^2$, $\lambda_{ij}$ is the tensor of thermal conductivity, $\eta$ is the proportionality coefficient having the dimension of s/K, $T$ is the absolute temperature, and $x_i$ are the Cartesian coordinates; summation is performed over repeated indices. Based on Eq. (2), an assumption of existence of the thermocurrent $\mathbf{j}^{te}$ in conducting LCs was put forward; this current is determined by the expression

$$\mathbf{j}^{te} \sim \nabla F_{te}, \quad j_i^{te} \sim \frac{\partial}{\partial x_i}\left(\frac{\partial T}{\partial x_k} \gamma_{kj} \frac{\partial T}{\partial x_j}\right). \tag{3}$$

If there is an electric field in the medium, the total current density with allowance for Eq. (3) can be written in the nonlinear form as

$$j_i = \sigma_{ij}\left[E_j - \alpha_{ij}\frac{\partial T}{\partial x_j} - \frac{\beta_{il}}{2}\frac{\partial}{\partial x_i}\left(\frac{\partial T}{\partial x_k}\gamma_{kj}\frac{\partial T}{\partial x_j}\right)\right]. \tag{4}$$

This equation involves a new thermoelectric tensor $\beta_{ij}$ (with the dimension of m$^3$/A) and, for generality, the conventional Seebeck thermocouple effect determined by the coefficient $\alpha_{ij}$; $\sigma_{ij}$ is the anisotropic conductivity of the medium. Charges possessing different mobilities (e.g., holes and electrons) are expected to contribute to conductivity. If the medium contacts some other materials on its boundaries, the thermoelectric coefficients in Eq. (4) depend on both contacting media.

Special experiments were performed to check the assumption of possible existence of the thermal current determined by the quadratic dependence (3); some of these experiments are described above.

As is seen from Eq. (4), the sign and magnitude of the thermocurrent can change only because of the linear Seebeck effect by virtue of the quadratic character of Eq. (3). If the heated and cooled sides of the converter are reversed ($T_1 \rightarrow T_2$, $T_2 \rightarrow T_1$), then the contribution of the thermocouple effect can be estimated from the difference in the measured thermovoltages. In all conducted experiments, the thermocouple voltage did not exceed 0.1 mV/K.

The proposed phenomenological dependence cannot yet be considered as a finalized theoretical model; nevertheless, as it follows from the experimental data reported above, the figure of merit of the thermoelectric effect predicted by Eq. (3) is comparable or even greater than that of other known TE conversion types. The maximum mean value predicted by Eq. (4) for the experimental data shown in Fig. 2 is $\beta\gamma \sim 5 \cdot 10^{-11}$ V·m$^2$/K$^2$ at 63ºC ($V \sim \beta\gamma(\Delta T/\Delta x)^2$). Equation (4) for the electric current ignores the diffusion current $j_i^D$, the effect of entrainment of electrons by phonons [3] proportional to the gradient of the concentration of the charge carriers

$$j_i^D = -D_{ij}^p \frac{\partial \rho^p}{\partial x_j} - D_{ij}^e \frac{\partial \rho^e}{\partial x_j} \tag{5}$$

($\rho^p$ and $\rho^e$ are the hole and electron charge density), and possible galvanic processes determined by the gradient of the chemical potential $\mu$ (in the general case, the values of $\mu$ for electrons and holes differ by the value of the energy gap between the electron and hole zones), which have the form

$$j_i^G = -G_{ij}\frac{\partial\mu}{\partial x_j}. \tag{6}$$

In writing Eq. (4), we also neglected various polarization, hydrodynamic [10], thermomechanical [14], and other phenomena, which are either typical for isotropic liquids or specific for LCs. There is no complete thermodynamic analysis in the paper (such a detailed analysis was performed long ago for the conventional Seebeck TE conversion [3, 15]); therefore, we do not consider the Peltier, Thomson, and Bridgeman effects for anisotropic media [3].

The tensor quantities involved into Eq. (4) for the current density in a uniaxial non-polar medium are given by the expressions

$$
\begin{aligned}
\lambda_{ij} &= \lambda_\perp \delta_{ij} + \lambda_a n_i n_j, & \lambda_a &= \lambda_\parallel - \lambda_\perp, \\
\sigma_{ij} &= \sigma_\perp \delta_{ij} + \sigma_a n_i n_j, & \sigma_a &= \sigma_\parallel - \sigma_\perp, \\
\alpha_{ij} &= \alpha_\perp \delta_{ij} + \alpha_a n_i n_j, & \alpha_a &= \alpha_\parallel - \alpha_\perp, \\
\beta_{ij} &= \beta_\perp \delta_{ij} + \beta_a n_i n_j, & \beta_a &= \beta_\parallel - \beta_\perp, \\
\gamma_{ij} &= \gamma_\perp \delta_{ij} + \gamma_a n_i n_j, & \gamma_a &= \gamma_\parallel - \gamma_\perp; \\
i, j &= 1, 2, 3 = x, y, z,
\end{aligned}
\tag{7}
$$

where the unit vector $\mathbf{n}$ indicates the crystal axis direction (for LCs, $\mathbf{n} = \mathbf{n}(t, x, y, z)$ is a director) and $\delta_{ij}$ is the Kronecker delta. The quantities indicated by the parallel and perpendicular subscripts are independent of the crystal axis direction $\mathbf{n}$, but are functions of thermodynamic variables. In LCs, which are representatives of soft matters [7, 9], many coefficients display a strong dependence on temperature; as is seen from experimental data (see Fig. 2), the additionally introduced coefficients $\beta$ and $\gamma$ are not exceptions from this rule, because it follow from Eq. (4) that $V \sim \beta\gamma$.

In constructing a thermodynamically consistent model, it will be necessary to take into account all factors affecting the current flow and energy release (absorption) in anisotropic liquids. The primary question to be answered is whether the densities and chemical potentials of the medium and charge carriers depend on the direction of the crystal axis $\mathbf{n}$ or of the director for LCs. Determination of $\rho = \rho\left(..., n_k, \ \partial n_i/\partial x_j\right)$, $\mu = \mu\left(..., n_k, \ \partial n_i/\partial x_j\right)$, where the ellipsis is understood as the dependence on other thermodynamic variables, is actually a problem of constructing the equation of state for anisotropic semiconducting liquids.

The TE converters described by Eq. (3) can be considered as specific devices for anisotropic liquids, because this effect (as well as the Benedicks effect) for isotropic liquid media and solid crystals is most probably very small. This can be demonstrated by a simple example of a one-dimensional isotropic rod with the ends heated to different temperatures and connected by a conductor made of another material. The thermocurrent for the material with weakly temperature-dependent parameters is found from Eq. (4) at $\alpha = 0$ as

$$j = -\frac{\sigma\beta\gamma}{2}\frac{\partial}{\partial x}\left(\frac{\partial T}{\partial x}\right)^2 = -\sigma\beta\gamma\frac{\partial T}{\partial x}\frac{\partial^2 T}{\partial x^2}. \tag{8}$$

The current for the connecting conductor is written in a similar manner. In Eq. (8), $x$ is the coordinate directed along the conductor. In writing Eq. (8), we ignore the electric field inside the conductors and the dependence of $\sigma\beta\gamma$ on temperature; therefore, the conductors outside the connections are assumed to be spatially homogeneous. The temperature distribution in the medium with the mass density $\rho$ and specific heat $c_p$ is described by the heat conduction equation, which has the following form if heat release due to current passage is ignored:

$$\rho c_p \frac{\partial T}{\partial t} \approx \lambda \frac{\partial^2 T}{\partial x^2}. \tag{9}$$

It follows from Eq. (9) that the thermocurrent (8) in the steady case is close to zero.

For an anisotropic TE conversion, Eq. (4) yields a rather complicated dependence of the thermocurrent on the temperature dynamics. Second derivatives with respect to the spatial variables appear in Eqs. (3) and (4), which can be expressed via the first derivatives of temperature with respect to time. The temperature distribution in an anisotropic medium is described by the equation of the form

$$\rho c_p \frac{\partial T}{\partial t} = \frac{\partial}{\partial x_i}\left(\lambda_{ij}\frac{\partial T}{\partial x_j}\right) + Q\,, \qquad (10)$$

where $Q$ is the energy density determined by the thermal processes in the medium. In the case with NLCs, the temperature and electric fields determine the direction (orientation) of the director $\mathbf{n}$ involved into the definition of the tensors (7). If the electrohydrodynamic processes are ignored, a simplified form of this dependence is given by the equation of the form [7, 10,13]

$$\Gamma\frac{\partial\mathbf{n}}{\partial t} = K\Delta\mathbf{n} + \frac{\varepsilon_a}{4\pi}\mathbf{E}\left(\mathbf{En}\right) + \gamma_a\nabla T\left(\nabla T\right) + \mathbf{G}\,, \qquad (11)$$

where $\Gamma$ is the parameter of orientation viscosity, $K$ is the elastic constant, $\varepsilon_a$ is the dielectric anisotropy, and $\mathbf{E}$ is the electric field vector. The vector $\mathbf{G}$ includes normalization parameters that ensure $(\mathbf{nn}) = 1$ and some effects that can be left unidentified for now and can be neglected.

Using the nonlinear equations (10) and (11), one cannot find an analytical dependence of the thermocurrent (4) on all other parameters introduced in Eq. (7) and the dynamic of the thermal energy conversion process as a whole. Moreover, there are many unclear issues in the influence of the LC – solid conductor or semiconductor interfaces. The interfaces between the media determine the LC director orientation and the galvanic processes determining charge exchange between the media [11].

Some qualitative conclusions can be deduced from analyzing the model (4), (10), (11). The constant direction of thermocurrent regardless of the sign of the temperature difference is a consequence of Eq. (2). The existence of the thermal orientation and thermomechanical effects [13,14] and the Fréedericksz effect (the electric field term in (11)) will lead, at increasing the heat flux and emergence of the electric field, to distortions of the original orientation structure of LC and can limit the thermocurrent. A characteristic temperature difference for emergence of the orientational distortions in LC has been determined in [13]. It is the same as in the present investigation, $\Delta T_{опт} \sim 1°$. For electric fields, the characteristic quantity of the Fréedericksz reorientation is a voltage of $\sim 1$ V [7,10]. The influence of these factors explains the experimentally obtained optimum value of the temperature difference.

To conclude, we can say that the TE conversion characteristics are not yet adequately systematized, but the results obtained up to now suggest that this effect can be used to create inexpensive alternative sources of energy.

# 4. CONCLUSIONS

The paper describes new experimental data on TE conversion in doped nematic liquid crystals. The discovered effect is not described by the classical Seebeck TE conversion. Theoretical justification of this effect and some other substances and regimes of the examined TE effect will be considered in subsequent studies. The characteristics reported in this paper do not exceed the maximum parameters obtained in semiconductor ($Bi_2Te_3$) TE converters based on the Seebeck effect, where $(ZT)_{BiTe} \approx 1$. A preliminary analysis and experimental data show that this effect can be used in the future to create a TE converter with $(ZT)_{LC} \approx 1$ and with the efficiency close to the efficiency of the Carnot cycle. The cost of one kilogram of bismuth telluride is approximately \$1000. In our estimates, the cost of substances necessary for the above-described TE converter (presumably, organic substances will be mainly used) will be appreciably lower.

The results obtained at the first stage of activities give us grounds to assume the prospects of using semiconducting anisotropic media in the development of alternative energy engineering.

This work was supported by the Interdisciplinary Integration Project No. 129 of the Siberian Branch of the Russian Academy of Sciences.